\documentclass[fleqn,10pt]{wlscirep}
\usepackage[utf8]{inputenc}
\usepackage[T1]{fontenc}
\usepackage{float}
\usepackage{chemformula}
\usepackage{graphicx}
\graphicspath{ {images/} }
\usepackage[export]{adjustbox}
\usepackage{wrapfig}
\usepackage{subcaption}
\usepackage[table,xcdraw]{xcolor}

\title{Computationally guided modifications of \textit{Cvi}UPO to improve catalytic activity}

\author[1]{Hanna-Friederike Poggemann}
\author[2]{Tim Dirks}
\author[1,3,4,*]{Timo Jacob}
\author[1,*]{Christoph Jung}
\affil[1]{Institute of Electrochemistry, Ulm University, Germany}
\affil[2]{Applied Microbiology, Faculty of Biology and Biotechnology, Ruhr University Bochum, Germany}
\affil[3]{Karlsruhe Institute of Technology (KIT), Germany}
\affil[4]{Helmholtz Institute Ulm (HIU) Electrochemical Energy Storage, Germany}

\affil[*]{timo.jacob@uni-ulm.de, karsten.jung@uni-ulm.de}


\begin{abstract}
Unspecific peroxygenases (UPOs) are promising biocatalysts that selectively oxyfunctionalize saturated hydrocarbons using only hydrogen peroxide as a co-substrate. However, peroxide-induced enzyme inactivation makes targeted enzyme engineering essential to mitigate this effect and additionally enhance catalytic performance. To meet this need, various systematic approaches are used, such as extensive database studies for rational enzyme design, as well as computational enzyme engineering. In this study, we followed the latter strategy and explored the possibility for computationally-guided modification of UPOs. Specifically, our focus was on uncovering the influence of active site amino acids on the catalytic activity of the enzyme \textit{Cvi}UPO. Two mutations were introduced close to the active center, and the changes in the energy barriers leading to the activated complex were investigated in detail by Quantum Mechanics/Molecular Mechanics Nudged Elastic Band (QM/MM-NEB) simulations. Our studies revealed that a change of the glutamic acid, assisting the catalytic cycle, by the shorter aspartic acid, leads to an increased reaction barrier, probably decreasing the catalytic activity of the enzyme. Exchanging the heme-anchoring cysteine group by a histidine, exhibited promising behavior as the energy barriers decreased significantly. However, it is possible that the histidine modification also alters the reaction behavior of the peroxygenase, turning it into a peroxidase, an aspect that so far could not be confirmed beyond doubt. Generally, simulations alone cannot conclusively determine whether substrate specificity and reactivity are maintained in both of the modifications tested. Nevertheless, our results highlight the importance of active pocket hydration and spin state for the catalytic reaction and demonstrate why a synergistic approach connecting theoretical predictions with experimental verifications is required for an efficient enzyme engineering.
\end{abstract}
\begin{document}
\flushbottom
\maketitle
\thispagestyle{empty}

\begin{figure}[ht]
\centering
\includegraphics[width=0.75\linewidth]{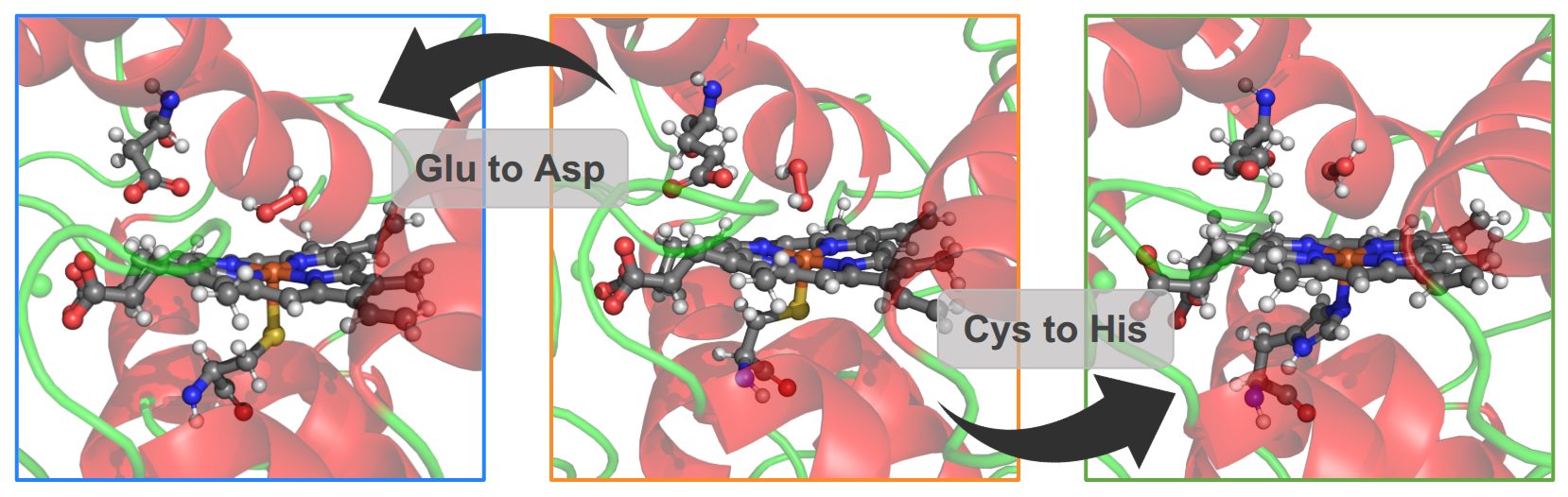}
\caption{Graphical abstract}
\label{fig:graphical_abstract}
\end{figure}

\newpage
\section*{Introduction} 
Biocatalysis, \textit{i.e.} catalysis using enzymes as natural catalysts, has gained a reputation as sustainable alternative to conventional catalysis. \cite{Wu2021BiocatalysisEnzymaticSynthesis} Enzymes are highly effective biocatalysts recognized for their remarkable substrate selectivity. Key characteristics such as their enantio- and stereoselectivity, high thermal stability compared to other proteins, and promising turnover rates make them a sustainable alternative to conventional catalysts. \cite{Wu2021BiocatalysisEnzymaticSynthesis} In this context, unspecific peroxygenases (UPOs) have been the focus of several studies in the past years, because they catalyze a diverse set of oxyfunctionalization reactions. \cite{Hofrichter2014Oxidationscatalyzedfungal, Karich2013Benzeneoxygenationoxidation ,Karich2016Exploringcatalaseactivity} UPOs are heme-thiolate enzymes only relying on the heme as cofactor and \ch{H2O2} as co-substrate without requiring any further potentially expensive additives, unlike most monooxygenases, making them a more cost-effective alternative. Furthermore, UPOs possess the interesting ability to hydroxylate unactivated hydrocarbons in a single catalytic cycle with minimal waste production -- an additional water molecule and the inactivated enzyme. \cite{Burek2019Hydrogenperoxidedriven} This distinguishes them from the peroxidases, which on the one hand are only able to cleave weak \ch{C-H} bonds and on the other hand simply perform a hydrogen abstraction subsequently releasing the substrate radical. \cite{Burek2019Hydrogenperoxidedriven, Hrycay2012monooxygenaseperoxidaseperoxygenase} However, several limitations must be addressed for this system to become a viable contender for industrial applications. To this point, only a few UPOs were discovered and studied in detail, and their natural function \textit{in fungi} is poorly understood. \cite{Hofrichter2022PeroxideMediatedOxygenation} In addition, they suffer from hydrogen peroxide induced inactivation when \ch{H2O2} is present in excessive concentrations. 
\cite{Karich2016Exploringcatalaseactivity} This leads to a dilemma as the \ch{H2O2} is needed for the catalytic reaction but may also be fatal to the enzyme. Consequently, either systems must be developed in which the dosage of \ch{H2O2} is specifically tuned to the system requirements, or the enzymes must be adapted through targeted modifications to prevent inactivation. \cite{Burek2019Hydrogenperoxidedriven} Ideally, these modifications are made to oxidation-prone amino acids in the active pocket near the heme center, as the inactivation mechanism most likely involves the formation of hydroxyl radicals that oxidize these amino acids and trigger a chain reaction leading to the destruction of the enzyme. \cite{Karich2016Exploringcatalaseactivity} 
However, amino acids in the active pocket have a strong influence on the enzyme's catalytic cycle, and the individual impact is poorly understood. 
\begin{figure}[H]
    \centering
    \includegraphics[width=0.7\linewidth]{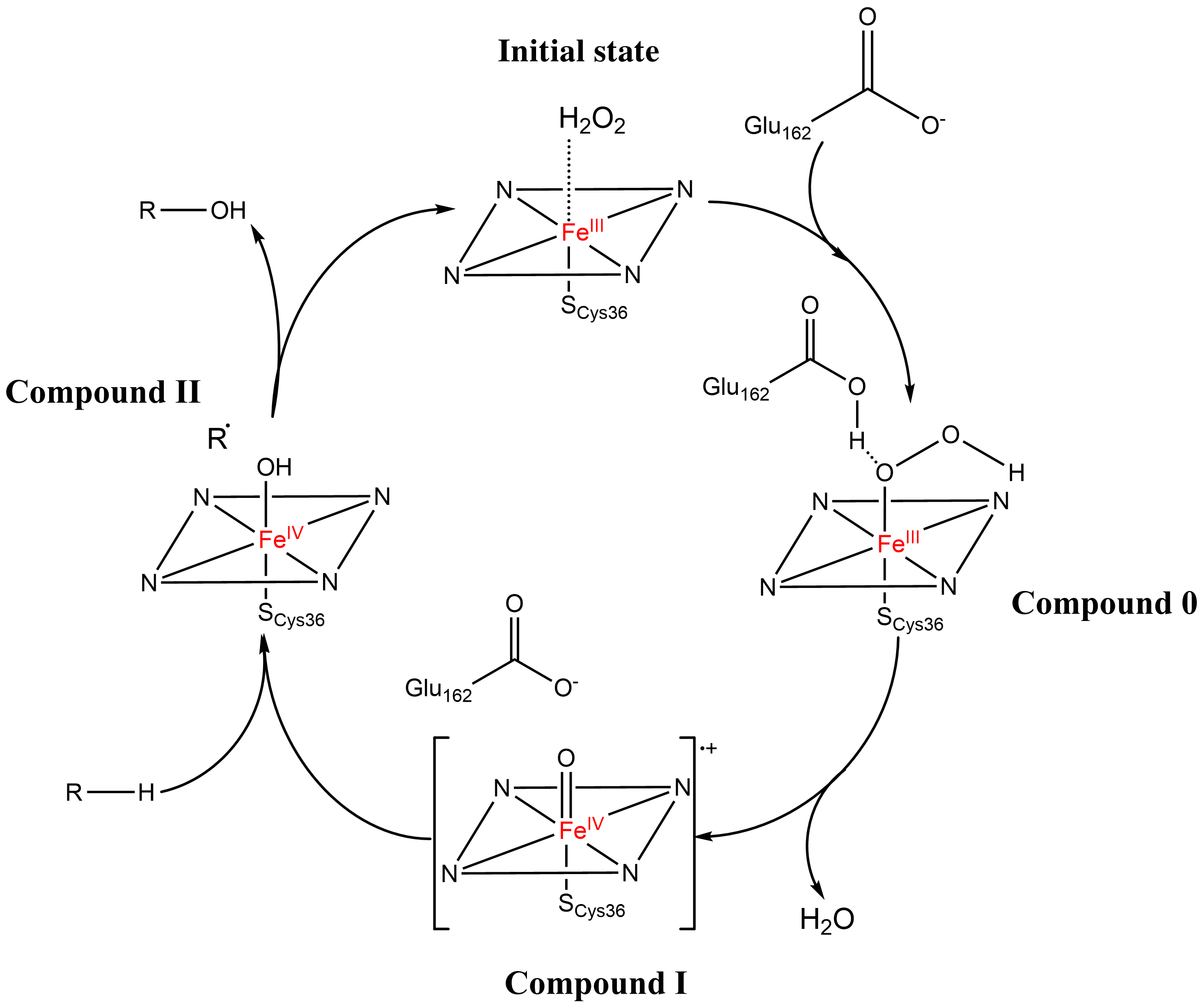}
    \caption[]{Catalytic cycle of \textit{Cvi}UPO following the heterolysis-homolysis mechanis. For simplicity, the cycle does not start with a water molecule as a distant ligand in the initial state but with the \ch{H2O2} already approaching the iron center. The second step is the formation of Compound 0, where the glutamic acid Glu162 is involved in abstracting one hydrogen from the \ch{H2O2}. Subsequently, this hydrogen recombines with the remaining \ch{OH} to form Compound I and an additional water molecule, this happens via a transition state that is not depicted here. Compound II evolves into Compound II by abstracting a hydrogen atom from the substrate. The resulting substrate radical then recombines with the hydroxyl group and leaves the active site, which is then back in its initial state.}
    \label{fig:Cat_cycle}
\end{figure}
Therefore, our study aims at elucidating the influence of active center amino acids in UPOs on the catalytic oxidation reaction, which will provide important insights to possibly enhance the enzyme’s stability and catalytic efficiency. As a model protein, we have focused on \textit{Cvi}UPO. The ascomycete \textit{Collariella virescens} (r\textit{Cvi}UPO) was recently discovered by sequence optimization of UPO genes and expressed in \textit{Escherichia coli} as active soluble enzyme. \cite{Linde2022StructuralCharacterizationTwo} 
Similar to the well-studied enzyme \textit{Aae}UPO, \textit{Cvi}UPO is a heme-thiolate enzyme naturally using hydrogen peroxide as co-substrate to catalyze the hydroxylation of the primary substrate. However, \textit{Cvi}UPO is a lot smaller than \textit{Aae}UPO (26–27\,kDa \textit{vs.} 46\,kDa) and can be expressed without glycosylations, which makes experimental analysis easier. However, the active pocket construction is quite similar to \textit{Aae}UPO, under the heme center there is the anchoring cysteine group (Cys19 in \textit{Cvi}UPO, Cys36 in \textit{Aae}UPO), and above a glutamic acid (Glu162 in \textit{Cvi}UPO, Glu196 in \textit{Aae}UPO). \cite{RamirezEscudero2018StructuralInsightsSubstrate, Linde2022StructuralCharacterizationTwo} 
Consequently, this enzyme most likely follows the same catalytic cycle as \textit{Aae}UPO, a combination of peroxide “shunt” pathway of P450-type enzymes and the typical cycle of these peroxidases. \cite{Sigmund2020Currentstatefuture, Hofrichter2022PeroxideMediatedOxygenation} This catalytic cycle consists of four main steps, schematically shown in Figure \ref{fig:Cat_cycle} for \textit{Cvi}UPO. It starts in the initial state or resting state with a distal water molecule attached to the heme center of the enzyme. This water molecule gets displaced by the \ch{H2O2}, which binds to the iron of the heme and subsequently looses one hydrogen to the Glu162, ultimately forming Compound 0, the ferric hydroperoxo complex (\ch{Fe-OOH}).
In the next step of the cycle, Compound 0 converts to Compound I, the iron–oxo complex, via a rearrangement of the \ch{OH-} on the \ch{Fe-OOH}. It recombines with the hydrogen atom on the Glu162 and splits off a newly formed water molecule, resulting in \ch{Fe=O} and \ch{H2O}. Compound I is the reactive state of the enzyme that is able to interact with the substrate to perform the catalytic reaction. Therefore, reaching Compound I energetically efficient is crucial for the catalytic performance of an enzyme. The \ch{Fe=O} complex abstracts a hydrogen atom from the substrate, thereby forming Compound II (\ch{Fe-OH}) and leaving behind a substrate radical. In the next step, the substrate radical recombines with the hydroxide group to form the hydroxylated substrate, whereupon the substrate is released from the active center, and the system returns to the initial state. 
A recent computational study by Costa \textit{\textit{et al.}} investigated the Compound I formation with \textit{Aae}UPO via two different mechanisms, the pure homolysis mechanism, involving a radical splitting of the \ch{H2O2} at the active center of the enzyme, and the heterolysis--homolysis or "Poulos--Kraut mechanism". They found the pure hemolysis mechanism to be significantly less favorable than the heterolysis-homolysis. \cite{Costa2023UnderstandingMultifacetedMechanism} Therefore, in our study we will concentrate on the heterolysis--homolysis mechanism that is sketched in Figure \ref{fig:Cat_cycle} for \textit{Cvi}UPO. \\
To explore how the molecular structure of the protein affects energy barriers along the pathway upon Compound I formation, we introduced mutations at two central amino acid positions in the active pocket and modeled the reaction pathway using a Quantum Mechanics/Molecular Mechanics (QM/MM) hybrid approach. The first modification, exchanging the glutamic acid Glu162 involved in the catalytic cycle for an aspartic acid, was chosen because we assumed that more space above the heme center and in the active pocket could improve the catalytic activity. Secondly, the anchoring cysteine amino acid Cys19 was replaced by a histidine, as we wanted to investigate whether the exchange of cysteine and histidine could increase the enzymes resistance against heme bleaching by the \ch{H2O2} or if the exchange would result in the peroxygenase adopting the reactive behavior of a peroxidase. This change is mechanistically plausible, because the cysteine ligand beneath the heme center plays a key role in breaking the substrate C–H bonds and facilitates the rebound step during substrate hydroxylation. The coordination of the cysteine sulfur to the heme iron donates additional electrons, increasing the electron density at the metal center and pushing it upwards, thereby facilitating hydrogen abstraction from the substrate. \cite{Groves2014Usingpushget, Green2004OxoironIVChloroperoxidaseCompound, Yosca2013IronIVhydroxidepKaRole, Wang2012DetectionKineticCharacterization, Wang2013DrivingForceOxygen} This effect has been uncovered and studied in detail by Green \textit{et al.} \cite{Green2004OxoironIVChloroperoxidaseCompound, Yosca2013IronIVhydroxidepKaRole} and Wang \textit{et al.}. \cite{Wang2012DetectionKineticCharacterization, Wang2013DrivingForceOxygen}
But to our knowledge it has not been investigated yet whether replacing the cysteine group impairs or completely prevents this ability.
The QM/MM method was employed in this study because this methods allow the investigation of the reactivity at the catalytic center at a QM level of theory while still including the effect of the surrounding protein at the MM level. The focus of the investigation was set on the first half of the catalytic cycle until the formation of Compound I, as it is the reactive key intermediate. Primarily, we conducted MD simulations to assess the stability of the enzyme mutants in a dynamic environment. This was followed by a QM/MM study of the important intermediates, namely the initial state, Compound 0, Compound I, and their electronic properties. Building on these results we performed Nudged Elastic Band \cite{Henkelman2000climbingimagenudged, Henkelman2000Improvedtangentestimate, Sheppard2011Pathswhichnudged} (NEB) simulations to identify the barriers along the reactive pathways for the native and the modified \textit{Cvi}UPOs. Finally, our substrate, ethylbenzene, was positioned in the active pocket to investigate the enzyme's affinity for it by MD simulations and QM/MM calculations of the main compounds as well as relaxed energy surface scans. 
We hope this theory-driven workflow will inspire new strategic approaches to enzyme engineering based on a deeper understanding of not only the steric but also electrostatic properties of the active center amino acids.

\section*{Results}
\subsection*{Characterization of the dynamic behavior} 
\begin{figure}[H]
    \centering
    \includegraphics[width=0.7\linewidth]{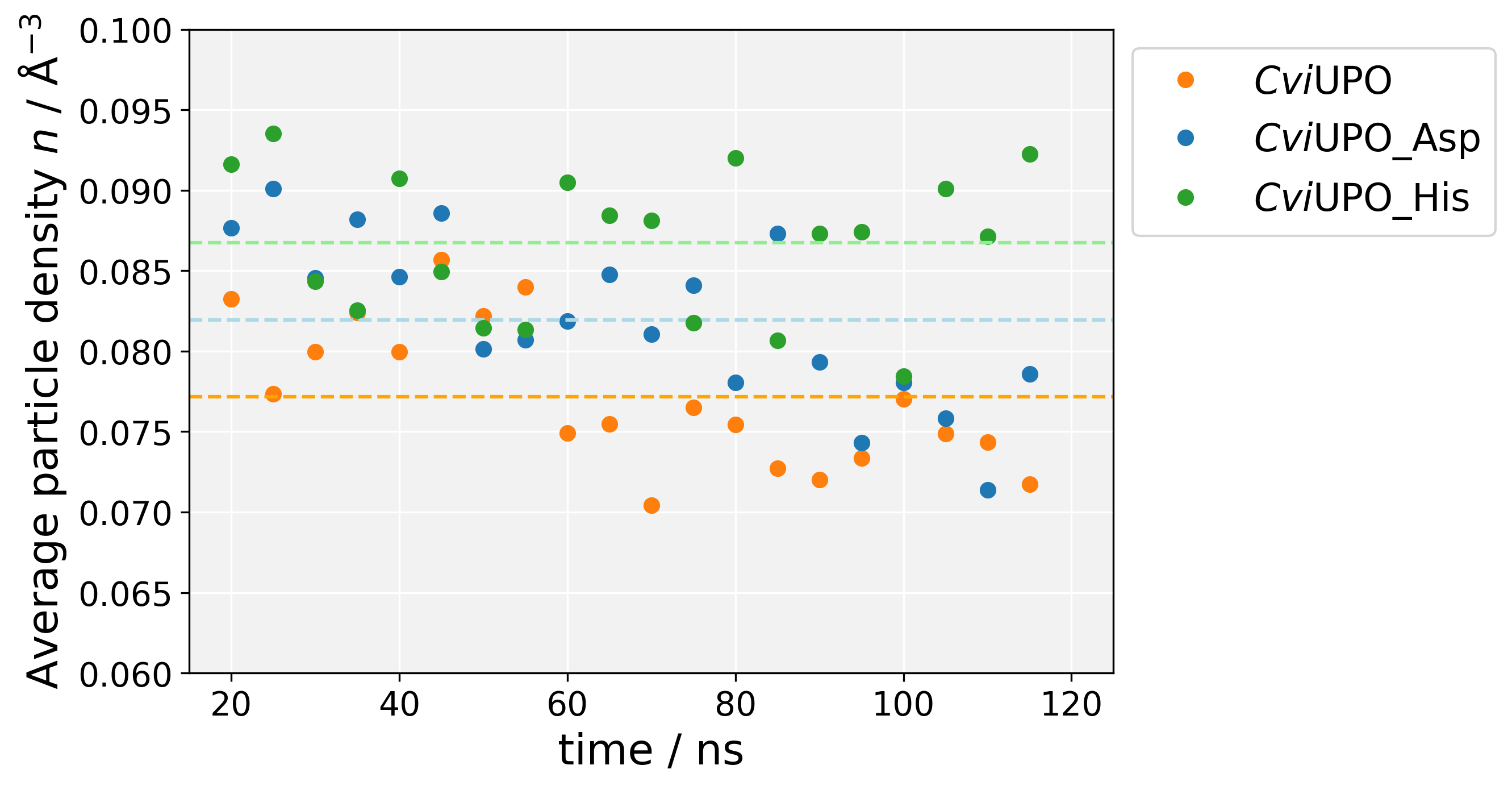}
    \caption[]{The plot shows the temporal progression of the average water molecule density in the active pocket for each of the enzymes, the native \textit{Cvi}UPO (orange), the \textit{Cvi}UPO with Asp modification (blue) and the \textit{Cvi}UPO with His modification (green), respectively. The dashed lines correspond to the respective time averaged values. For the analysis 120\,ns MD simulations were performed -- the first 20 were excluded from the plot -- and for each data point, the average value was calculated over 5\,ns.}
    \label{fig:Cvi_MD_results}
\end{figure}

The differences in dynamic behavior of the three enzymes -- the native \textit{Cvi}UPO, the aspartic acid modified \textit{Cvi}UPO (Asp \textit{Cvi}UPO), and the histidine modified \textit{Cvi}UPO (His \textit{Cvi}UPO) -- were initially studied with classical MD simulations. All enzymes were stable during the total simulation time of 120\,ns and did not show any unexpected sign of unfolding at 300\,K. The radius of gyration rG is similar between the proteins, only His \textit{Cvi}UPO is slightly smaller than the other two enzymes with rG being 1.713\,nm in contrast to 1.730\,nm and 1.739\,nm for native \textit{Cvi}UPO and Asp \textit{Cvi}UPO, respectively. To gain deeper insights into the modifications that had an impact on the active pocket itself, the volume of the active pocket was analyzed over time, as well as the hydration of the pocket from those values an average particle density of water molecules inside the active pocket was calculated. The  average water molecule density is shown in Figure \ref{fig:Cvi_MD_results}, the pocket volumes and the pocket hydration can be found in Figure S1 in the SI . Even though the native \textit{Cvi}UPO (orange markers) has the biggest pocket volume (Figure S1 (a)) the amount of water molecules inside the pocket are in a similar range than for the other two enzymes, resulting in the lowest total time averaged water molecule density, in comparison. However, the pocket volume increased after 50 ns, while the number of water molecules within the pocket remained relatively constant throughout the simulation, resulting in a downward trend in the average water molecule density. 
This behavior is to be expected, as the system moves dynamically and each data point represents an average value over 5\,ns. Notably, the Asp \textit{Cvi}UPO (dashed blue line) shows a higher average water molecule density than the native \textit{Cvi}UPO, due to a lower average pocket volume and a similar amount of hydration inside the pocket. This contrasts with prior assumptions, as the introduction of Asp placed a shorter amino acid in the active pocket, which was expected to increase pocket volume rather than reduce it. 
The His \textit{Cvi}UPO shows the highest average water molecule density in comparison. Still the average hydration in the pocket stays similar to the other two enzymes but the average volume is lower for this enzyme. Partially, the amino acid residues Leu64 and Ile61 are responsible for the smaller active pocket as they reach deeper into the pocket in the case of the His modification compared to the other two structures. It must be mentioned, also, that the temporal variation where rather high for the His \textit{Cvi}UPO and the molecules pocket seemed to be less stable, which could be a side effect from the modification. Replacing the cysteine Cys19 at the heme center with histidine appears to result in a larger restructuring of the enzyme than the modification of the glutamic acid Glu162.\\ 
In summary, it can be said that all enzymes exhibit very similar pocket hydration despite their differing pocket volumes. This is generally a good prerequisite for the catalytic efficiency of enzymes, as a lower water content in active pocket might diminish the catalytic activity of this enzyme since studies. In particular, the work of Taher \textit{et al.}\cite{Taher2023Computationallyguidedbioengineering} has shown that the water in the active pocket plays a key role in catalytic efficiency and improving the active pocket hydration can lead to higher catalytic activity in enzymes. Furthermore, the active pocket hydration also seems to play an important role for a fast product release. \cite{Kim2025Fastproductrelease}
\subsection*{Comparison of Compound II formation of native and modified \textit{Cvi}UPO}
Molecular dynamics reach their limitations when it comes to the investigation of reactivity and electronic properties of a system. Therefore, QM/MM simulations were conducted to elucidate the reactive pathway of the catalytic reaction for the three variations of \textit{Cvi}UPO. The focus here was on the formation of Compound I, the reactive iron–oxo complex, that reacts with the substrate of choice. A reduced reaction barrier between initial state and Compound I could indeed have a positive effect on the catalytic activity. As already mentioned in the introduction \textit{Cvi}UPO uses \ch{H2O2} as a co-substrate and follows generally the catalytic cycle that was proposed for \textit{Aae}UPO. \cite{Costa2023UnderstandingMultifacetedMechanism} The QM/MM simulations were conducted in all three spin states the iron atom at the center of the heme can adopt, as spin crossovers are entirely possible during the reaction. The low spin (LS) state refers to the state in which, if possible, all valence electrons in the system are arranged in pairs. In the intermediate spin (IS) state half of the valence electrons are paired and the other half is unpaired, while in the high spin state (HS) all valence electrons try to be unpaired. For a clear schematic illustration, please refer to the publication by Mariusz Radoń. \cite{Radon2014SpinStateEnergetics} Naturally, the histidine-modified \textit{Cvi}UPO exhibits a different spin state, as the total number of electrons differs from the native and the aspartic acid modified \textit{Cvi}UPO, in which cysteine serves as the anchoring amino acid. 
\begin{figure}[H]
\centering
\includegraphics[width=0.8\linewidth]{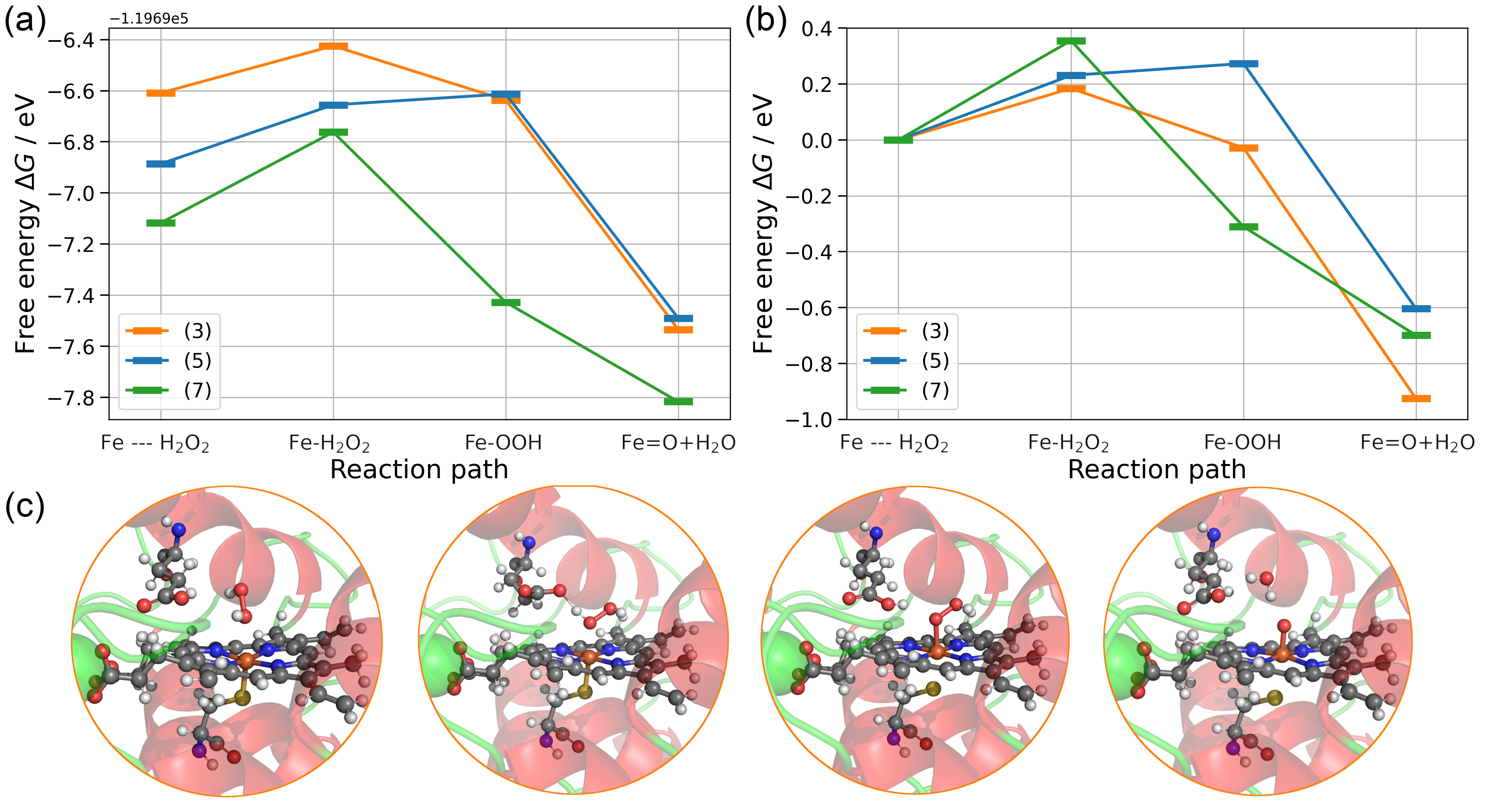}
\caption{Free energy diagram of native \textit{Cvi}UPO. The $x$-axis shows the state of the iron co-substrate complex in which the system is located, while the $y$-axis shows the free energy $G$ in (a) and $\Delta G$ in (b), respectively. In (b), the values of each curve are referenced to the initial state. Each spin state is represented by its own color, the LS is shown in orange, the IS in blue, and the HS in green. The models shown in panel (c) display a 3D-representation of the active center corresponding to the four states of the system plotted on the $x$-axis of each graph.}
\label{fig:Cvi_wt_dG}
\end{figure}
\begin{figure}[H]
    \centering
    \includegraphics[width=0.9\linewidth]{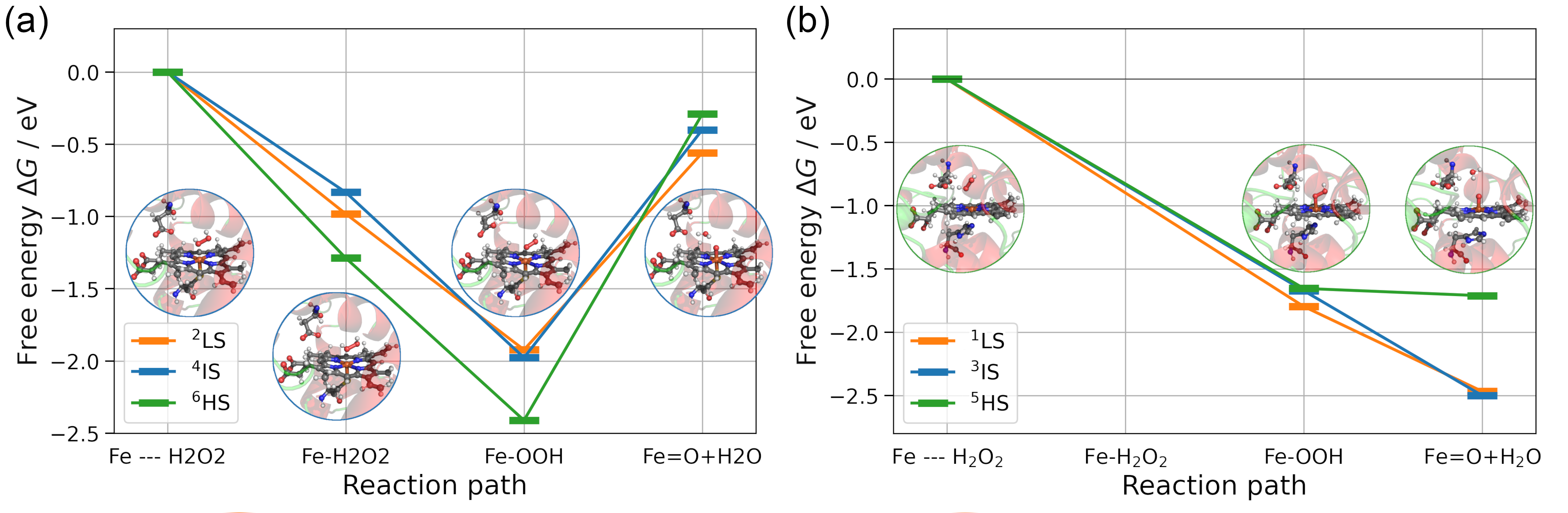}
    \caption[]{The plots show the free energy diagram of Asp-modified \textit{Cvi}UPO in (a) and His-modified \textit{Cvi}UPO in (b). Each curve is referenced to the initial state. Each spin state is represented by its own color, the LS is shown in orange, the IS in blue, and the HS in green. The corresponding inlets display the 3D-representations of the active center in the different states.}
    \label{fig:Cvi_asp_his_dG}
\end{figure}
Four important steps along the formation of Compound I were identified for closer investigation. The initial state with \ch{H2O2} as distal ligand and not yet bound to the iron atom in the catalytic center, the intermediate state where \ch{H2O2} is weakly bound to the iron, Compound 0 with the \ch{Fe-OOH} conformation and the additional hydrogen bound to Glu162 and finally Compound I, the ferro-oxide state with a newly formed \ch{H2O} molecule (\ch{Fe=O + H2O}).
Figure \ref{fig:Cvi_wt_dG} shows the free energy diagram of these four states, not normalized in (a) and normalized to the first state in (b), the respective 3D representations of the active center in each state are shown underneath in (c). Even though all spin states are close in energy, Figure \ref{fig:Cvi_wt_dG} (a) reveals that the high spin state has slightly lower energies, most likely making it more favorable than the other two spin states. However, the normalized values in Figure \ref{fig:Cvi_wt_dG} (b) highlight that the free energy difference between the initial state \ch{Fe---H2O2} and the \ch{Fe-H2O2} intermediate state is higher in the HS compared to the other spin configurations. With respect to the energy differences, the LS pathway looks most favorable, because the energy difference between the \ch{Fe---H2O2} and the \ch{Fe-H2O2} state is only 0.2 eV. However, we cannot draw a final conclusion from these simulations, as no energy barriers between the points are taken into account.\\
Figure \ref{fig:Cvi_asp_his_dG} shows the free energy diagram for the Asp-modified \textit{Cvi}UPO in \ref{fig:Cvi_asp_his_dG} (a) and His-modified \textit{Cvi}UPO in \ref{fig:Cvi_asp_his_dG} (b). Both differ significantly from the free energy profile of native \textit{Cvi}UPO. The energy profile of Asp-modified \textit{Cvi}UPO (Figure \ref{fig:Cvi_asp_his_dG}) (a) exhibits a strong minimum at Compound 0, consequently making Compound I energetically less favorable than Compound 0. This behavior is not beneficial for the catalytic reaction, as the system would never reach Compound I, and the reaction would stop at in Compound 0. Interestingly, this holds for all spin states. Additionally, the not normalized free energy values, which can be found Figure S2 in the SI, are nearly identical in energy in all spin states for this enzyme.\\
The free energy diagram for His \textit{Cvi}UPO in Figure \ref{fig:Cvi_asp_his_dG} (b) has another distinctive feature, the \ch{Fe-H2O2} intermediate state is not stable. During geometry optimization the system directly relaxes into Compound 0, therefore  the \ch{Fe-H2O2} intermediate state is not shown in Figure \ref{fig:Cvi_asp_his_dG}. This result indicates that there is no energy barrier between the initial state where \ch{H2O2} is not bound and Compound 0 for the His-modified \textit{Cvi}UPO, this could proof beneficial for the catalytic activity. In contrast to Asp \textit{Cvi}UPO the final state Compound I is lower in energy than Compound 0, making this important state of the catalytic cycle energetically more favorable than Compound 0. Similar to Asp \textit{Cvi}UPO, though, the different spin states are all very close in energy, enabling the possibility for spin crossing during reaction. However, the results of spin-unrestricted DFT calculations should always be interpreted with care, even if the spin contamination is low. Furthermore, the applied exchange--correlation functional B3LYP was found to favor low lying intermediate spin states and slightly underestimate the stability of low spin states. \cite{Schoeneboom2004RestingStateP450cam, Radon2014SpinStateEnergetics} The comparison of the free energy diagrams leads to the assumption that His \textit{Cvi}UPO is probably the more favorable modification of the enzyme than Asp \textit{Cvi}UPO and could even be catalytically more active than the unmodified \textit{Cvi}UPO. Nevertheless, the energy barriers in between the stages of the catalytic cycle must be examined in detail in order to make a definitive statement.

\begin{figure}[H]
    \centering
    \begin{subfigure}{0.5\textwidth}
    \includegraphics[width=1\textwidth]{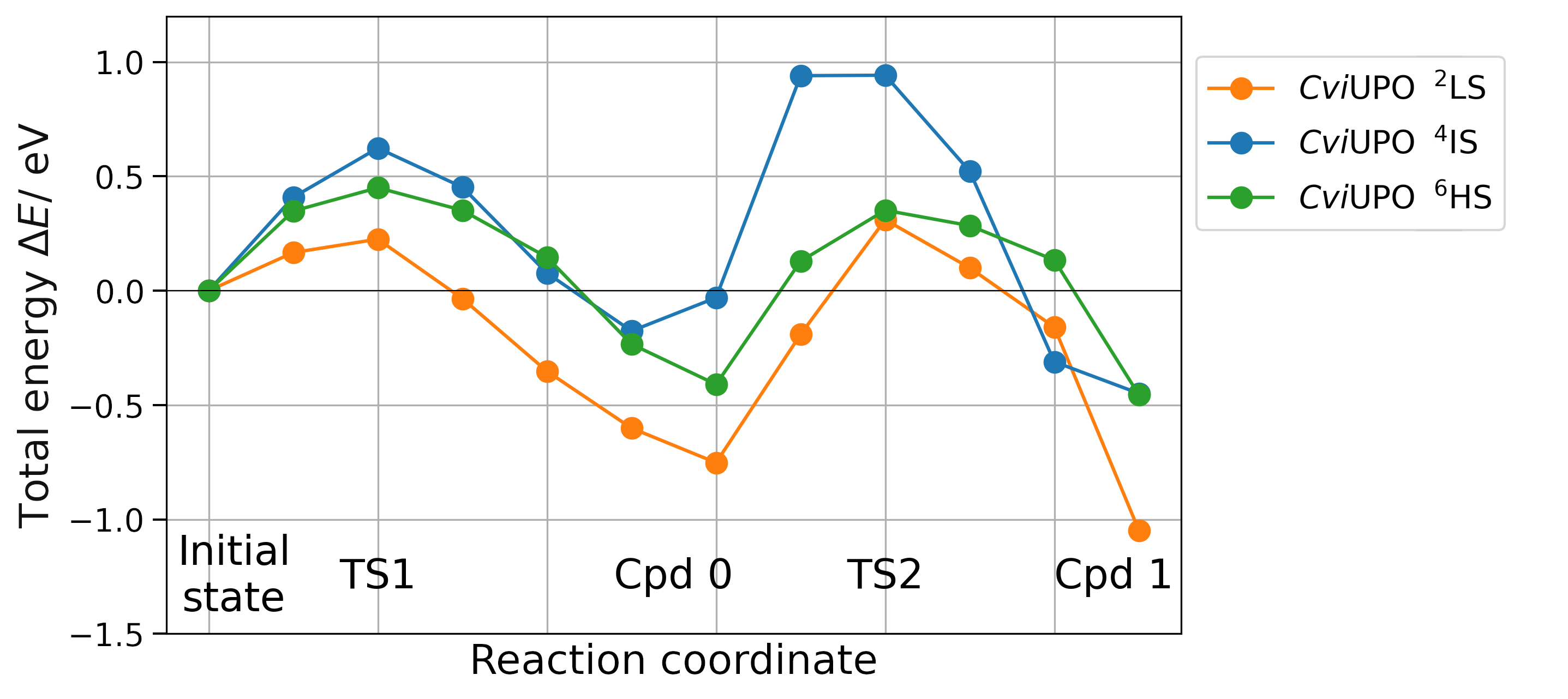}
    \caption{}
    \label{fig:Cvi_wt_NEB}
    \end{subfigure}
    \\[1ex]
    \begin{subfigure}{0.5\textwidth}
    \includegraphics[width=1\textwidth]{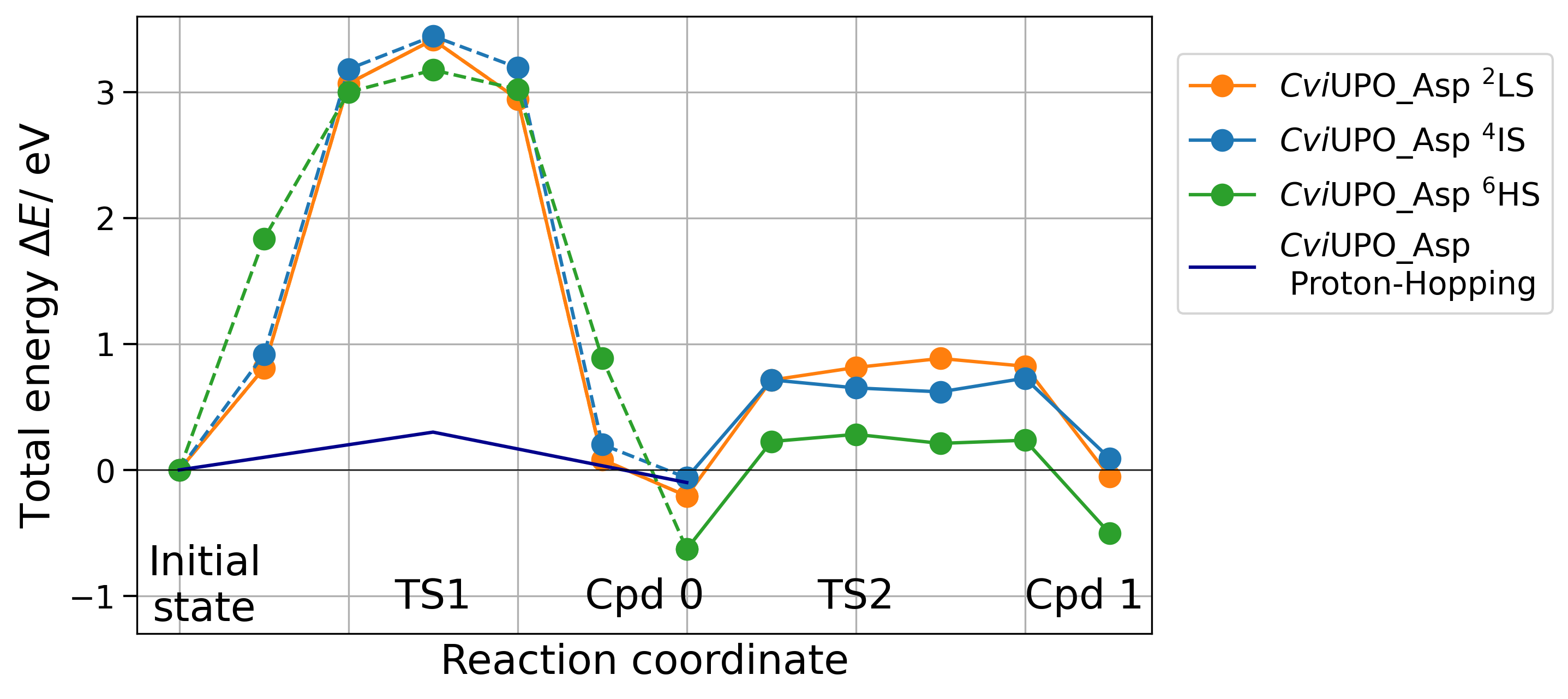}
    \caption{}
    \label{fig:Cvi_asp_NEB}
    \end{subfigure}
    \\[1ex]
    \begin{subfigure}{0.5\textwidth}
    \includegraphics[width=1\textwidth]{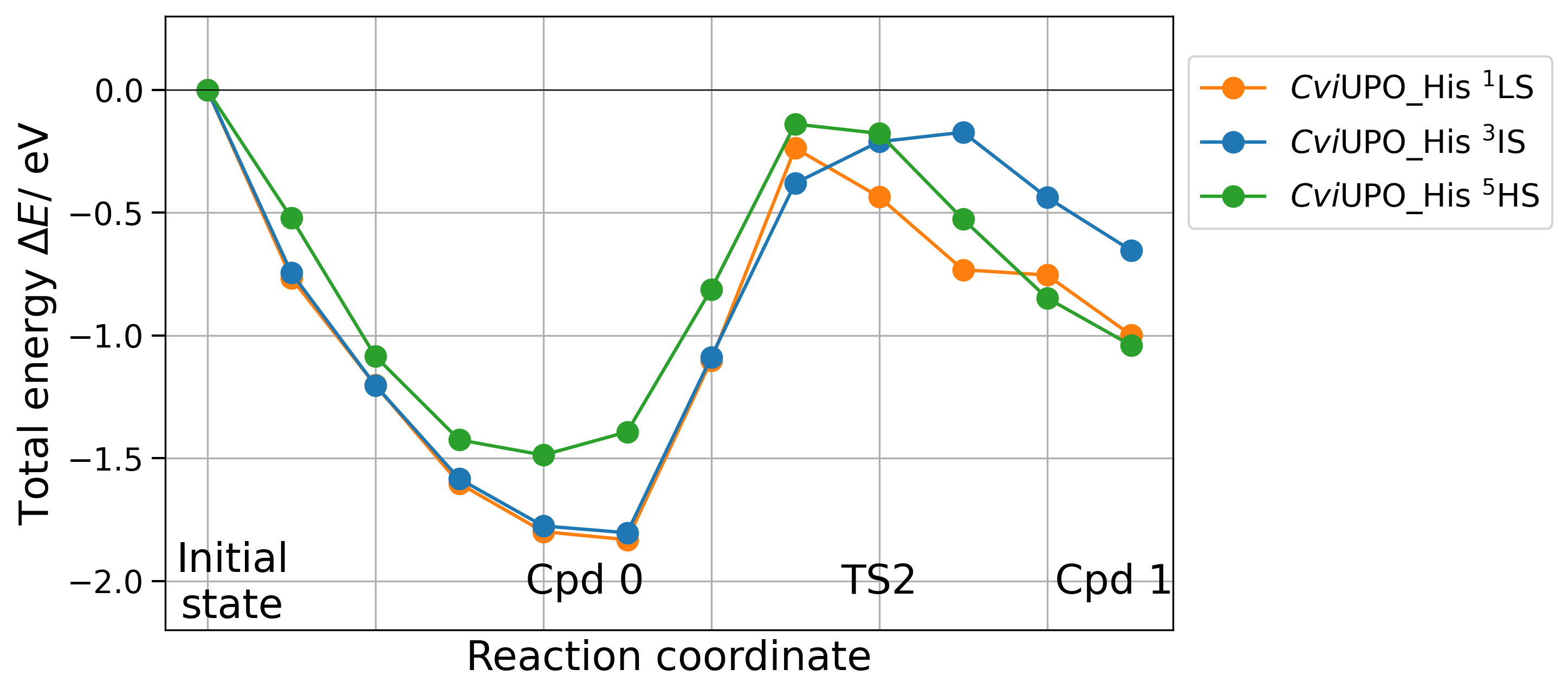}
    \caption{}
    \label{fig:Cvi_his_NEB}
    \end{subfigure}
    \caption[]{Reaction pathway from the initial state of the catalytic cycle via Compound 0 (Cpd 0 in the plot) towards Compound I (Cpd 1 in the plot) for native \textit{Cvi}UPO (a), \textit{Cvi}UPO with aspartic acid modification (b), and \textit{Cvi}UPO with histidine modification (c), derived by NEB calculations. The values of each curve are referenced to its initial state. Each spin state is represented by its own color, the LS is shown in orange, the IS in blue, and the HS in green. The respective values for the reaction barriers can be found in Table \ref{tab:Cvi_dE_barriers}.}
    \label{fig:Cvi_NEB}
\end{figure}

Therefore, in the next step NEB simulations were conducted for each enzyme system in all three spin configurations. The results are displayed in Figure \ref{fig:Cvi_NEB} for each of the three enzyme systems, respectively. The previously optimized key states along the catalytic cycle were employed as start and end states for the NEBs. Figure \ref{fig:Cvi_wt_NEB} shows that the calculated pathway of the native \textit{Cvi}UPO exhibits two energy barriers. The first barrier is located between the \ch{Fe---H2O2} initial state, and the local minimum of the curve in Compound 0. In the maximum of the curve, the transition state TS1, lies the intermediate state with the \ch{Fe-H2O2} complex. The second barrier lies between Compound 0 and Compound I with TS2 at the maximum of the curve corresponding to a \ch{Fe-O-OH-H-Glu162} complex, where the \ch{OH} has not yet completely detached from the ferro-oxide complex and is on the verge of recombining with the hydrogen that migrated to Glu162 in Compound 0 (see Figure \ref{fig:Cat_cycle} step 2 and 3). The second barrier is higher than the first one (the exact values of all barriers can be found in Table \ref{tab:Cvi_dE_barriers}) and might therefore be more difficult to overcome. Even if barriers of approximately 1\,eV seem rather high, the obtained values are in the same range as energy barriers reported for other enzymes such as Chloroperoxidase and \textit{Aae}UPO. \cite{Chen2008QuantumMechanical/MolecularMechanical, Costa2023UnderstandingMultifacetedMechanism}
\begin{table}[H]
\centering
\caption{Energy barriers from the NEB calculations for all three enzymes. For native \textit{Cvi}UPO and \textit{Cvi}UPO with aspartic acid modification the values of the fist and the second barrier, corresponding to the maximum at TS1 and TS2 respectively, are listed for all spin states. As the \textit{Cvi}UPO with histidine modification exhibited only one maximum in the second half of the NEB, merely the values for the second barrier are listed. The values for the second barrier are calculated in reference to the previous local minimum (LM). All barriers were calculated for the forward direction of the catalytic cycle.}
\begin{tabular}{cc|c|c|}
\cline{3-4}
\multicolumn{1}{l}{}                                                             & \multicolumn{1}{l|}{}      & \cellcolor[HTML]{EFEFEF}\textbf{first barrier / eV} & \cellcolor[HTML]{EFEFEF}\textbf{second barrier / eV} \\ \hline
\multicolumn{1}{|c|}{\cellcolor[HTML]{EFEFEF}\textbf{}}                          & \cellcolor[HTML]{EFEFEF}$\mathrm{{}^2LS}$ & 0.225                                               & 1.062                                                \\
\multicolumn{1}{|c|}{\cellcolor[HTML]{EFEFEF}\textbf{native \textit{Cvi}UPO}} & \cellcolor[HTML]{EFEFEF}$\mathrm{{}^4IS}$ & 0.624                                               & 1.118                                                \\
\multicolumn{1}{|l|}{\cellcolor[HTML]{EFEFEF}}                                   & \cellcolor[HTML]{EFEFEF}$\mathrm{{}^6HS}$ & 0.450                                               & 0.762                                                \\ \hline
\multicolumn{1}{|c|}{\cellcolor[HTML]{EFEFEF}\textbf{}}                          & \cellcolor[HTML]{EFEFEF}$\mathrm{{}^2LS}$ & 3.414                                               & 1.092                                                 \\
\multicolumn{1}{|c|}{\cellcolor[HTML]{EFEFEF}\textbf{Asp \textit{Cvi}UPO}}       & \cellcolor[HTML]{EFEFEF}$\mathrm{{}^4IS}$ & 3.443                                               & 0.778                                                \\
\multicolumn{1}{|l|}{\cellcolor[HTML]{EFEFEF}}                                   & \cellcolor[HTML]{EFEFEF}$\mathrm{{}^6HS}$ & 3.174                                               & 0.910                                                \\ \hline
\multicolumn{1}{|c|}{\cellcolor[HTML]{EFEFEF}\textbf{}}                          & \cellcolor[HTML]{EFEFEF}$\mathrm{{}^1LS}$ & -                                                   & 1.594                                                \\
\multicolumn{1}{|c|}{\cellcolor[HTML]{EFEFEF}\textbf{His \textit{Cvi}UPO}}       & \cellcolor[HTML]{EFEFEF}$\mathrm{{}^3IS}$ & -                                                   & 1.631                                                \\
\multicolumn{1}{|l|}{\cellcolor[HTML]{EFEFEF}}                                   & \cellcolor[HTML]{EFEFEF}$\mathrm{{}^5HS}$ & -                                                   & 1.253                                                \\ \hline
\end{tabular}
\label{tab:Cvi_dE_barriers}
\end{table}
Of course, the presence of a substrate in the active pocket, as well as the water molecules, which are not included in the QM region for these calculations, might influence the barriers. While the presence of water is generally thought to support or mediate the catalytic reaction, \cite{Taher2023Computationallyguidedbioengineering, Teune2025Insightswatermediated} and therefore properly decreases the energy barriers, the substrate in the pocket might even increase them, at least this behavior was reported for \textit{Aae}UPO by Costa \textit{et al.}. \cite{Costa2023UnderstandingMultifacetedMechanism} As already indicated in the free energy diagrams, the intermediate spin state is lowest in energy for the first half of the scanned reaction pathway (Figure S3 (a) in the SI). After Compound 0, the energies of the LS and the HS are overlapping, hence a spin crossover might be possible at that point. This possibility is supported by the fact that the HS displays the lowest energy difference between Compound 0 and TS2. \\
Furthermore, until reaching TS2 the HS has the lowest energy values. In TS2, though, the IS and the HS are overlapping and leading from TS2 to Compound I, the IS exhibits the lowest energy values again. Although it must be emphasized that, for the second half of the NEB all three spin states are close in energy and present energy barriers that are thermodynamically surmountable, for example, by to the presence of water or the involvement of other amino acids. In contrast to the presented results, the computational studies by Chen \textit{et al.} and Costa \textit{et al.} found the doublet spin state (LS) to be the most favorable during the catalytic reaction for the investigated enzymes cytochrome $\mathrm{P450_{cam}}$ and \textit{Aae}UPO, respectively. \cite{Chen2008QuantumMechanical/MolecularMechanical, Costa2023UnderstandingMultifacetedMechanism} However, neither study was conducted using the B3LYP functional, and it must be taken into account that, as mentioned above, B3LYP overstabilizes the IS.\cite{Schoeneboom2004RestingStateP450cam, Radon2014SpinStateEnergetics} Furthermore, the experimental results by Wang \textit{et al.} showed that \textit{Aae}UPO undergoes a  LS to HS interconversion upon substrate binding. \cite{Wang2012DetectionKineticCharacterization} Therefore, the spin transitions calculated here are well within the range of physically plausible phenomena.
The NEB calculation along the reactive pathway of Asp \textit{Cvi}UPO, Figure \ref{fig:Cvi_asp_NEB}, also displays two energy barriers. However, the maximum of the first barrier, TS1, does not correspond to the \ch{Fe-H2O2} state but the transition from the \ch{Fe-H2O2} to Compound 0 in the local minimum of the curve. The reason for this enormous barrier of approximately 3\,eV for all spin states is that the hydrogen atom has to cross about three Angstrom of free vacuum space to get to Asp162, as Asp162 is shorter than Glu162. This process should be viewed purely in qualitative terms, which is why it is shown only as a dotted line in Figure \ref{fig:Cvi_asp_NEB}, though, as the hydrogen atom would normally move to Asp162 along the water molecules in the active pocket via a proton hopping mechanism. This effect could not be covered by the simulation, but a hypothetical barrier (calculated with 0.1\,eV per hopping step \cite{Elahi2025DFTassessmentactivation}) is indicated in the plot in the blue line. However, the hight of the second barrier between Compound 0 and Compound I is in a similar range as for the native \textit{Cvi}UPO, even though the OH released from the active center has to again move about two Angstrom across free space to recombine with the hydrogen on Asp162. Of course, interactions with solvent molecules might also decrease the barrier here, but the effect is more difficult to estimate, therefore it is not illustrated in Figure \ref{fig:Cvi_asp_NEB}. As already evident by the free energy plot, the spin multiplicities result in nearly the same energy for the initial state, Compound 0 and Compound I of the investigated pathway, but the curve progression differs slightly. The LS and the IS are almost identical in energy for the fist half of the NEB, while HS is slightly higher in energy (see Figure S3 (b) in the SI). Nevertheless, the HS has the lowest energy barrier, as the difference between the initial state and TS1 is slightly smaller than for the other spin states. For the second half of the NEB all three spin states are nearly identical in terms of their energy values (Figure S3 (b)), as are the barriers (see table \ref{tab:Cvi_dE_barriers}). \\
Looking at Figure \ref{fig:Cvi_his_NEB} it is evident that the His modified \textit{Cvi}UPO has only one energy barrier the system has to overcome during the reaction. The free energy diagrams gave the right incitation predicting that there would not be a barrier between the initial state and Compound 0. The barrier in the second half of the NEB, between Compound 0 and Compound I, varies little between the spin multiplicities, but it is lowest for the HS with 1.25\,eV. Again, the spin states are close in energy and cross in several points, indicating spin flip possibilities. However, contradicting to the free energy diagram (Figure \ref{fig:Cvi_asp_his_dG} (b)), the final state, Compound I, is not lower in energy than Compound 0. This result suggests that the catalytic cycle could be trapped in the local minimum at Compound I since, on the one hand, it has to overcome a barrier that is approximately 0.5 eV higher than for the other enzymes and, on the other hand, there is no favorable low energy state on the other side of the barrier. However, it cannot be ruled out that the NEB has only converged to a local minimum in its last point, rather than the lowest energy minimum. 

\subsection*{Influence of the substrate}
\begin{figure}[H]
\centering
\includegraphics[width=1\linewidth]{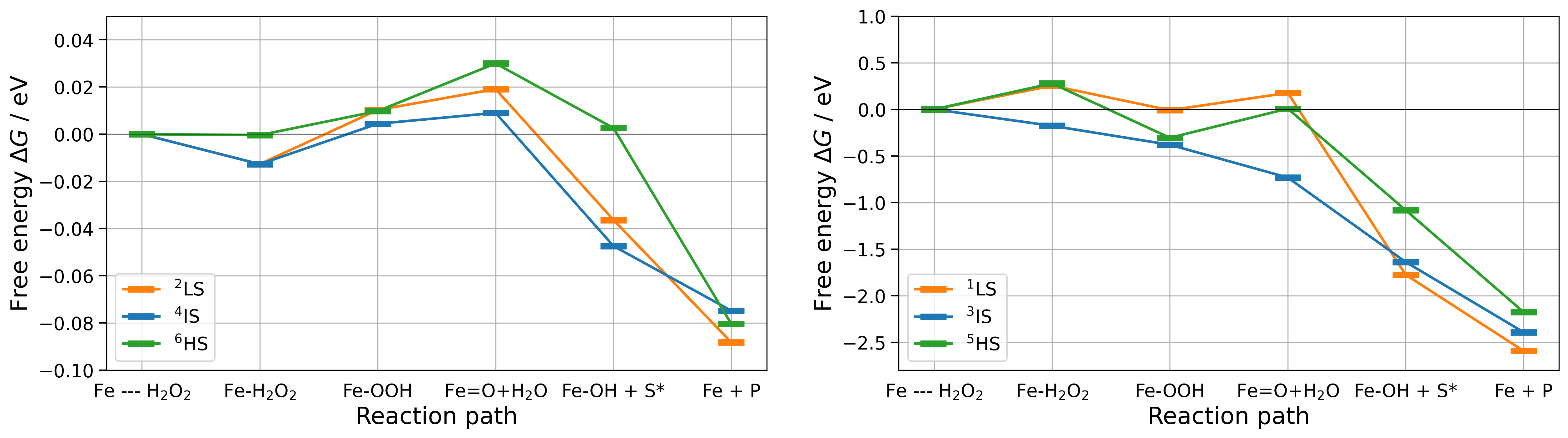}
\caption{Free energy diagram of native \textit{Cvi}UPO (a) and His \textit{Cvi}UPO (b) with ETBE in the active pocket. Each spin state is represented by its own color, the LS is shown in orange, the IS in blue, and the HS in green. The x-axis shows the state of the iron co-substrate complex in which the system is located and the $\mathrm{\Delta G}$ is plotted on the y-axis. Each curve is referenced to its initial state, the non-normalized values can be found in Figure S4 (a) and (b) in the SI.}
\label{fig:Cvi_wt_his_ebte_dG}
\end{figure}

The reaction pathway towards Compound I indicated that the histidine modification might be favorable for the catalytic reaction, while the Asp modification seems to be less suitable. However, not only does the formation of the reactive ferro-oxide complex influences the enzymatic activity, but the stability of the substrate in the active pocket also plays a dominant role. Therefore three different initial positions of ETBE in the active pocket of all three enzyme variants, derived from docking simulations, were tested for their stability in 100\,ns MD simulations. All initial structures with the ETBE in the various positions are provided alongside the publication. ETBE in the first position proved to be stable in the pocket for the native and for the histidine modified \textit{Cvi}UPO. The substrate does rotate throughout the simulation, but it does not move out of the pocket. In contrast, ETBE is not stable in the pocket of the \textit{Cvi}UPO with the aspartic acid modification. The second and third positions tested proved to be unstable positions for ETBE in the pocket of all three enzymes. 
This result underlines what was already insinuated by the study of the Compound I formation. The aspartic acid modification does not seem to be beneficial for the catalytic reactivity. A final assessment can only be made by experimentally investigating the enzyme activity, though. \\
To gain some deeper insights into whether the histidine modification outperforms the native enzyme, we again conducted QM/MM geometry optimizations of the key intermediates and additionally relaxed energy surface scans with the substrate ETBE in the active pocket for these two enzymes. Figure \ref{fig:Cvi_wt_his_ebte_dG} displays the free energy differences calculated for all states of the catalytic cycle -- the initial state, Compound 0, Compound I, Compound II, and the initial state with hydroxylated substrate, (R)-1-phenylethanol -- for the native \textit{Cvi}UPO (Figure \ref{fig:Cvi_wt_his_ebte_dG} (a)) and the histidine modified \textit{Cvi}UPO (Figure \ref{fig:Cvi_wt_his_ebte_dG} (b)).
For both enzymes the trend of the curves differs visibly from the free energy calculations without the substrate (Figure \ref{fig:Cvi_wt_dG}). The native \textit{Cvi}UPO exhibits a destabilization of Compound I, the ferro-oxide state, with respect to the Compound 0 and the \ch{Fe-H2O2} transition state independently of the chosen spin state. In the calculations without the substrate, Compound I had the lowest free energy value, in Figure \ref{fig:Cvi_wt_his_ebte_dG} (a) it has the highest energy. However, the last two points of the cycle then show a negative energy difference, with the product state as the clear minimum for all spin states. Probably there is an energy barrier between Compound I and Compound II, as well as between Compound II and the product state, which are not represented here, but will be investigated in the relaxed surface scans. The His \textit{Cvi}UPO also shows a destabilization of Compound I compared to the calculations without substrate in the pocket but only for the low and high spin state. In the intermediate spin state the energy is continuously decreasing towards the product state. For the LS and the HS the value with the highest energy difference is the transition state with the bound \ch{H2O2}. However, the product state has the lowest energy for all three spin states. Marking the state with the completed reaction as the most favorable one of the catalytic cycle. For both enzymes, the native \textit{Cvi}UPO and the His \textit{Cvi}UPO, the substrate occupancy of the active site seems to either destabilise Compound I or hinder the binding of the co-substrate \ch{H2O2}. This observation is consistent with results reported by Costa \textit{et al.} for \textit{Aae}UPO. \cite{Costa2023UnderstandingMultifacetedMechanism} The results from this study as well as those from Costa \textit{et al.} suggest that the energetically preferred mechanism involves initial binding of the co-substrate \ch{H2O2} and formation of Compound I prior to substrate entry into the active site. This reaction sequence has been experimentally validated by Wang \textit{et al.}, \cite{Wang2012DetectionKineticCharacterization} who demonstrated higher reaction rates for the enzyme \textit{Aae}APO when Compound I formation precedes substrate binding. \\
The relaxed surface scans were solely performed in the intermediate spin state, as the free energy diagrams indicated this might be the most favorable spin state. Therefore, the reaction coordinate of the Compound II formation -- removing the hydrogen (H13 in the structure file) from the substrate and binding to Compound I -- was scanned, as well as the transfer of the resulting hydroxyl group from the heme center to the substrate, forming the product (R)-1-phenylethanol. The energy barrier for the Compound II formation in the native \textit{Cvi}UPO is 1.37\,eV high while the histidine modified \textit{Cvi}UPO has a barrier of 0.67\,eV. Counterintuitively, the first barrier of the native \textit{Cvi}UPO is higher than for His \textit{Cvi}UPO and relatively high in general. This might be due to an unfavorable initial position of ETBE in the active pocket of the native \textit{Cvi}UPO. Furthermore, the barriers between both enzymes are not perfectly comparable, as the position of the substrate in the pocket differs between the two enzymes. The second step shows barriers of 0.89\,eV, and 0.48\,eV for the native \textit{Cvi}UPO and the histidine modified \textit{Cvi}UPO, respectively. However, it should be noted that with His \textit{Cvi}UPO, there is no rebound step of the hydroxyl group from the heme center to the substrate. Instead, the water molecule previously split off during the formation of Compound 1 binds to the substrate and transfers one of its hydrogen atoms to the \ch{Fe-OH} complex. This behavior indicated that the energy barrier for the rebound step might be too high to overcome for the histidine modified \textit{Cvi}UPO and explains the lower reaction barrier in this step compared to the native \textit{Cvi}UPO. In the relaxed surface scan of the native \textit{Cvi}UPO, the previously mentioned water molecule seems to play a critical role. It seems to support the rebound step of the hydroxyl group and remains as a distal water molecule over the heme center after the product has formed. \\
It must be emphasized, however, that scans of the relaxed energy surface are not as accurate as NEB calculations, and the actual transition states have probably not been pinpointed exactly.
Furthermore, neither the exact orientation of the substrate in the active pocket nor the expected transition states for the transition from Compound II to the product (R)-1-phenylethanol are known. This makes it difficult to accurately assess the quality of these simulations, which is why the results should be viewed critically. Interestingly, for both enzymes the transition Compound II to the hydroxylated substrate showed an involvement of the \ch{H2O} molecule, which was split off in the formation of Compound I. This suggests that adequate solvation of the active site is essential for efficient substrate conversion because it facilitates this specific step of the catalytic cycle. 
Despite these insights, the simulations did not reveal whether the altered electron density at the heme center, resulting from cysteine replacement by histidine, impedes the substrate deprotonation, and the result is not conclusive enough to prove beyond doubt that the one step hydroxylation of the substrate cannot be performed with the His modification. In particular, the simulations with the substrate in the active pocket have shown that an experimental investigation of the modified \textit{Cvi}UPO structures is necessary in order to clarify which modification is advantageous.


\section*{Discussion}
The MD simulations conducted for the native and the modified enzymes proved that the modified enzymes are stable over an extended period of simulation time. The detailed analysis of the active pockets showed a slightly decreased pocket volume for the Asp modified \textit{Cvi}UPO and a even more decreased pocket volume for the His modified \textit{Cvi}UPO. The hydration of the active pocket, however, is similar for all three enzymes. So, the MD simulations alone are no sufficient indicator if one of the conducted modifications is beneficial.  

The free energy calculations and subsequent NEB calculations, however, revealed that the Asp modification is most likely not an advantageous. From the NEB calculations we can conclude that Asp modification could hinder the deprotonation of the \ch{H2O2} because Asp is shorter than Glu, and the proton therefore has to overcome a greater distance to bind to the aspartic acid. As a result, the NEB calculations displayed high energy barriers for all three spin states in the first half of the catalytic cycle, especially between the initial state and Compound 0 (\ref{fig:Cvi_asp_NEB}). However, this result must be considered in light of the model limitations, like the missing water in the QM region, as explained above. It is possible that the process would be supported via proton hopping along the water molecules in the active pocket under experimental conditions. Consequently, this and the other results from the Asp modified \textit{Cvi}UPO should be interpreted qualitatively only.\\
At first glance, the His modified \textit{Cvi}UPO seems to be a promising modification. The free energy diagram suggested that there would be no barrier to overcome between the initial state ans Compound 0, which was confirmed by the NEB calculations. But the NEB also revealed that the second barrier, between Compound 0 and Compound I, is higher than in the other systems, which increases the chance that the catalytic reaction will stop in Compound 0. For this reason, it cannot be clearly determined whether the His modification exhibits an energetically improved catalytic pathway compared to the native \textit{Cvi}UPO. \\
There is no question that the spin state of the iron atom in the center of the heme decisively influences the catalytic cycle of heme proteins. However, if the simulation results presented here are examined in detail, it stands out that especially for the Asp modified \textit{Cvi}UPO and the His modified \textit{Cvi}UPO, the calculated energies are nearly identical for all spin states.  Therefore, no definitive statement can be made about the preferred spin state for either of the two enzyme modifications. This fact could potentially be related to the limitations of the DFT method with regard to the correct description of spin states -- especially in unrestricted open shell simulations. \cite{Radon2014SpinStateEnergetics, Jacob2012Spindensityfunctional} However, it could also indicate that even small changes in the local electronic environment of the heme center significantly influence the preferred electronic configuration of the iron center. This assumption is supported by the known spin degeneracy of heme enzymes, where the LS, IS, and HS states are often separated by only small energy differences. \cite{Li2012Recentdensityfunctional, Shaik2010P450EnzymesTheir, Radon2014SpinStateEnergetics} 
The behavior of the native \textit{Cvi}UPO differs slightly from the modified enzyme. The energies of the LS, IS, and HS are more clearly separated, and in the NEB results, mostly the IS shows the lowest energy values, closely followed by the LS, which is in line with the expectation, as the literature is emphasizing that the B3LYP destabilizes the LS with respect to a low-lying IS. \cite{Radon2014SpinStateEnergetics} \\
Although the MD simulations, including the substrate, confirmed that ETBE is stable in the active pocket of the His modified \textit{Cvi}UPO, the relaxed surface scans did not show clear results whether the second half of the catalytic cycle, starting from Compound I, is improved or worsened by the His modification in comparison to the native \textit{Cvi}UPO. Furthermore, there has been no evidence of a trend suggesting that the His modification hinders the hydroxylation of the substrate, even though the studies by Green and Wang would suggest this. \cite{Green2004OxoironIVChloroperoxidaseCompound, Yosca2013IronIVhydroxidepKaRole, Wang2012DetectionKineticCharacterization, Wang2013DrivingForceOxygen} However, there is little information about the exact nature of the transitions between Compound I and receiving the product (R)-1-phenylethanol with the heme back in its initial state, these simulations can only make rough assumptions that lack a profound basis. In order to make a definitive statement, the second half of the cycle for both the native \textit{Cvi}UPO and the His \textit{Cvi}UPO would have to be examined in greater detail and compared to experimentally determined structures. \\
This led to the conclusion that even though the simulations elucidated the fundamental processes of \textit{Cvi}UPO's catalytic cycle and highlighted the importance of the spin for enzymatic reaction chemists, the results are insufficient to fully assess whether the Asp modification or the His modification would increase the catalytic activity of \textit{Cvi}UPO. Neither do the simulations prove that replacing Cys and His can convert a peroxygenase into a peroxidase. In order to finally clarify these questions, both modifications would have to be investigated experimentally. Preliminary attempts to express the His modified \textit{Cvi}UPO have shown that this is not so easy to achieve in practice. Both the enzyme yield after purification and the subsequent heme loading were low. Further experiments will be necessary to determine whether this modification is unfavorable for the enzyme or whether the expression method needs to be improved accordingly. In any case, some studies have succeeded in expressing an enzyme, where the anchoring cysteine amino acid at the heme center was replaced by a histidine. \cite{McIntosh2015StructuralAdaptabilityFacilitates} \\
In summary, it is important to highlight that simulations and experiments are both equally important for informed enzyme engineering, which goes beyond simple trial and error. And although molecular dynamics simulations are sufficient for predicting simpler modifications, like the engineering of water channels, high-level quantum chemical calculations are indispensable to gain a deeper understanding of the electronic interactions that shape catalytic reactions at the enzyme's active center. We therefore consider a computationally guided workflow including quantum chemical calculations to be extremely valuable, particularly when tuning the reactions at the active center of an enzyme.

\section*{Methods}
\subsection*{Molecular Dynamics Simulations}
The structure of the enzyme \textit{Cvi}UPO was taken from the crystal structure of \textit{Cvi}UPO (PDB code: 7ZCL. \cite{2022UnspecificperoxygenaseCollariella}) The PDB file was cleaned of all additional residues not belonging to the enzyme, and hydrogen atoms were added to the protein structure using the H++ web-based protonation prediction tool (version 4.0), aiming for neutral pH. \cite{Gordon2005Hserverestimating, Myers2006simpleclusteringalgorithm, Anandakrishnan2012H3.0automating}
The resulting protein structure was solvated at neutral pH and equilibrated using the GROMACS simulation software suite (version 2022.2 released June 16th, 2022) \cite{2022GROMACS2022.2Source, Abraham2015GROMACSHighperformance, Berendsen1995GROMACSmessagepassing, Hess2008GROMACS4Algorithms, Lindahl2001GROMACS3.0package, Pall2015TacklingExascaleSoftware, Pall2020Heterogeneousparallelizationacceleration, Pronk2013GROMACS4.5high, VanDerSpoel2005GROMACSFastflexible} with the latest version of the CHARMM force field, CHARMM36, for GROMACS (Updated July 2022). \cite{SoterasGutierrez2016Parametrizationhalogenbonds, Vanommeslaeghe2010CHARMMgeneralforce, Vanommeslaeghe2012AutomationCHARMMGeneral, Vanommeslaeghe2012AutomationCHARMMGenerala, Yu2012ExtensionCHARMMgeneral} After the protein was solvated in a cubic simulation box with periodic boundary conditions (80x80x80 \AA), the equilibration steps were performed following the suggested procedure of J. A. Lemkul. \cite{Lemkul2019ProteinsPerturbedHamiltonians} Primarily, the system was energy minimized and followed by a short simulation of 100\,ps in a thermodynamic ensemble with a constant number of particles, constant volume and constant absolute temperature, $NVT$ ensemble, and subsequently a 100\,ps equilibration in a $NPT$ ensemble, with a constant number of particles, constant pressure and constant absolute temperature. All enzyme atoms but the hydrogen atoms where position restrained during these steps. Afterwards, a 10\,ns MD simulation at 300\,K was conducted for further equilibration without any restraints. From this simulation, a snapshot of the fully equilibrated \textit{Cvi}UPO system was chosen to apply the desired protein modification. The glutamic acid with the residue number 162 was replaced by an aspartic acid using the molecular visualization program PyMOL \cite{PyMOL} to create \textit{Cvi}UPO with the aspartic acid modification. Furthermore, the cysteine group underneath the heme center, Cys19, was replaced by histidine for the \textit{Cvi}UPO structure with the histidine modification. Both modified structures were again equilibrated for additional 10\,ns. Finally, the production MD simulations of all three systems, native \textit{Cvi}UPO, \textit{Cvi}UPO with Asp, and \textit{Cvi}UPO with His, were performed for a total duration of 100\,ns again at 300\,K. \\
The resulting trajectories of all systems were analyzed with respect to the active pocket volume and hydration of each enzyme. The pocket volume analyzer POVME2 (POcket Volume MEasurer 2) developed by the Durrant Lab \cite{Durrant2011POVMEAlgorithmMeasuring, Durrant2014POVME2.0Enhanced} was used for the analysis of the active pocket volume. Building on the resulting volume geometry, a Python script was developed relying on the Python API of the Ovito Open Visualization Tool \cite{Stukowski2009Visualizationanalysisatomistic} to calculate the number of water molecules in the pocket. The Python scripts for the pocket volume and hydration analysis are available on GitHub ( \url{https://github.com/hpoggemann/}). \\
For the MD simulations with the substrate in the active pocket of the enzymes, a snapshot from the respective simulations was chosen, and the solvent molecules were removed again. Subsequently, the molecular docking program AutoDock Vina \cite{Trott2010AutoDockVinaimproving, Eberhardt2021AutoDockVina1.2.0} was used to place the substrate, ethylbenzene (ETBE), in the active pocket of each protein, to generate an initial guess for the binding positions. Of the positions predicted by Vina, three different confirmations with the lowest energies were chosen for further investigation, resulting in nine structures. Afterwards, the proteins with substrate in the active pocket were then solvated and equilibrated again, and 100\,ns production runs were executed to determine the stability of each substrate positioning in the pocket of the enzyme variants. 

\subsection*{QM/MM simulations}
A Quantum Mechanics/Molecular Mechanics (QM/MM) simulation is a hybrid simulation combining a quantum mechanical (QM) calculation with an MD simulation. It generally consists of a small inner region, QM region, that models the chemically interesting part of a system by first-principles methods, like density functional theory (DFT), and a bigger outer region describing most of the system at MM level. Applying this simulation type, we are able to investigate the catalytic reaction at the reactive center of an enzyme while preserving the chemical environment of the rest of the protein and the solvent. \cite{Senn2006QM/MMMethodsBiological} \\
For the QM/MM simulations, a snapshot from the MD production run of each enzyme was chosen, and the solvent was removed. AutoDock Vina \cite{Trott2010AutoDockVinaimproving, Eberhardt2021AutoDockVina1.2.0} was used to place the \ch{H2O2} in the active pocket of each enzyme, and the enzymes were solvated with a sphere of water molecules. The equilibration routine was performed, and the final structures were used as input for the QM/MM simulations. \\
All QM/MM simulations were carried out with the quantum chemistry software ORCA. \cite{Neese2012ORCAprogramsystem, Neese2025SoftwareUpdateORCA} The CHARMM force field parameters were used for the MM part of the QM/MM simulation, and the QM region was described with the B3LYP functional, the triple-zeta basis set def2-TZVP with D4 dispersion correction. Even though there exist great alternative functionals at this point, the classical B3LYP functional was chosen because it is well characterized, especially regarding its behavior of over- and understabilization of certain spin states. \cite{Radon2014SpinStateEnergetics} The interactions between the QM and the MM region were described by an electrostatic embedding scheme, and the additive QM/MM approach was used. \cite{Neese2025ORCA6.0Manual} Most atoms of the input structure were part of the MM region and kept frozen during the QM/MM simulation. To reduce the computational cost, only the amino acids within an 8\,\AA~radius around the QM region were tagged as active atoms and therefore allowed to move during the simulations. The QM region contained the heme center of the enzyme, the anchoring cysteine group Cys19 or His19 in the case of the histidine modification, the glutamic acid Glu162 or Asp162 in the modified version, and the co-substrate \ch{H2O2}. The covalent bonds in the QM and the MM region were capped with the hydrogen link atom approach, automatically handled by ORCA. \cite{Neese2025ORCA6.0Manual} As different spin states were investigated, all simulations were calculated with unrestricted open shell. This system setup was used for the geometry and frequency optimizations as well as for the nudged elastic band (NEB) simulations of all three enzymes. \\
In order to generate the different states of the catalytic cycle of \textit{Cvi}UPO and its modified versions -- the initial state with distal \ch{H2O2}, the intermediate state with bound \ch{H2O2}, Compound 0 and Compound I -- the starting structure was modified accordingly. After optimization and frequency analysis -- in low, intermediate, and high spin state -- these states were used as starting and end points for the NEB calculations to elucidate the energy barriers along the catalytic cycle. 
The NEB was therefore divided into two parts to record the process leading from the initial state to Compound I in detail. The first half begins in the initial state, denote as reactant R in Figure \ref{fig:Cvi_NEB} (a) to (c) and ends at Compound 0, while the second part begins at Compound 0 and ends at Compound I, the product P of the NEB calculation. The climbing image NEB (CI-NEB) implementation of ORCA was used, employing the default convergence criterion and the free end option for additional optimization of the starting and end images. After convergence, the climbing image was optimized again using the transition state optimizer to ensure the true maximum was found.
In addition to the QM/MM calculations without substrate, we also conducted simulations with ETBE in the active pocket for the native \textit{Cvi}UPO and the histidine modified \textit{Cvi}UPO. The aspartic acid modified \textit{Cvi}UPO was not considered for simulations with ETBE, as the MD simulations revealed no stable position of the substrate in its pocket. Again, a snapshot from the previous MD simulations was chosen, and the system preparations were conducted as described above. The QM region contained again the heme center, the anchoring amino acid, Cys19 or His19, the glutamic acid Glu162, as well as the co-substrate \ch{H2O2} and the substrate ETBE. Subsequently, the starting structures of the key compounds were designed for both systems and optimized, including a frequency analysis. After optimization, relaxed energy surface scans were conducted to investigate the reaction pathway leading to Compound II and the substrate radical, followed by the substrate radical recombination with the hydroxyl group forming (R)-1-phenylethanol. 

\bibliography{Lit_Bio-Catalysis}

\section*{Acknowledgments (not compulsory)}

C.J. and T.J. gratefully acknowledge funding through the DFG (CRC 1316-2). The authors acknowledge support by the state of Baden-Württemberg through bwHPC and the DFG through grant no INST 40/575-1 FUGG (JUSTUS 2 cluster).

\section*{Author contributions statement}
H.-F. P. designed the study, performed the simulations as well as the respective data analysis and wrote the manuscript. T.D. conceived the preliminary experiment(s) and analyzed the experimental results. C.J. and T.J. provided the funding and supervised the work. All authors reviewed the manuscript.

\end{document}